\documentclass[12pt]{article}
\usepackage{arxiv}

\usepackage[utf8]{inputenc} 
\usepackage[T1]{fontenc}    
\usepackage{hyperref}       
\usepackage{url}            
\usepackage{booktabs, threeparttable}       
\usepackage{amsfonts, amssymb, amsmath, latexsym, amsthm, bm}       
\usepackage{nicefrac}       
\usepackage{microtype}      
\usepackage{lipsum}		
\usepackage{graphicx}
\usepackage{natbib}
\usepackage{doi}
\usepackage{pdflscape}
\usepackage{listings}
\title{Regularizing threshold priors with sparse response patterns in Bayesian factor analysis with categorical indicators}
\author{ \href{https://orcid.org/0000-0002-9114-3896}{\includegraphics[scale=0.06]{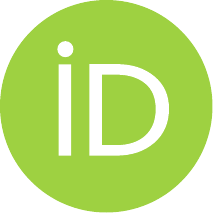}\hspace{1mm}R. Noah Padgett} \\
	Department of Epidemiology\\
	Harvard T.H. Chan School of Public Health University\\
	Boston, MA  \\
	\texttt{npadgett@hsph.harvard.edu} \\
	\And
	{Grant B. Morgan} \\
	Department of Educational Psychology\\
	Baylor University\\
	Waco, TX \\
	\texttt{grant\_morgan@baylor.edu} \\
	\And
	\href{https://orcid.org/0000-0001-9458-6185}{\includegraphics[scale=0.06]{orcid.pdf}\hspace{1mm}Tim Lomas} \\
	Human Flourishing Program\\
	Harvard Institute for Quantitative Social Science\\
	Cambridge, MA \\
	\texttt{tlomas@hsph.harvard.edu} \\
}

\renewcommand{\shorttitle}{Regularizing Threshold Priors}

\hypersetup{
pdftitle={A project evaluating prior specification for thresholds in Bayesian SEM},
pdfsubject={stat.ME},
pdfauthor={R. Noah Padgett, Grant B. Morgan, Tim Lomas},
pdfkeywords={Bayesian, thresholds, Dirichlet, induced-prior, SEM, factor analysis, item response theory},
}

\begin{document}
\large
\maketitle

\begin{abstract}
Using instruments comprising ordered responses to items is ubiquitous for studying many constructs of interest. However, using such an item response format may lead to items with response categories infrequently endorsed or unendorsed completely. In maximum likelihood estimation, this results in non-existing estimates for thresholds. This work focuses on a Bayesian estimation approach to counter this issue. The issue changes from the existence of an estimate to how to effectively construct threshold priors. The proposed prior specification reconceptualizes the threshold prior as prior on the probability of each response category, which is an easier metric to manipulate while maintaining the necessary ordering constraints on the thresholds. The resulting induced-prior is more communicable, and we demonstrate comparable statistical efficiency with existing threshold priors. Evidence is provided using a simulated data set, a Monte Carlo simulation study, and an example multi-group item-factor model analysis. All analyses demonstrate how at least a relatively informative threshold prior is necessary to avoid inefficient posterior sampling and increase confidence in the coverage rates of posterior credible intervals.
\end{abstract}


\section*{Introduction}

Data are messy,  as any data analyst will readily admit. 
A common contributing factor to messy data is the response format used to collect observations.
Psychological,  educational,  and patient-reported observation data are often ratings, graded responses, or categorical in nature.
Using discrete observations is often a straightforward way to obtain individuals' perceptions towards a topic \citep{DeVellis2016, Fowler2014, Fink2003}, but this approach also comes with many decisions about the number of, and labels for, response categories.
Including too few response categories may result in not collecting sufficient information to identify discernible differences among individuals, whereas including too many response categories may result in individuals being unable to discern differences in the categories \citep{Matell1971, Garner1951}.
Researchers commonly use between two and ten labeled,  ordered categories with five or seven being the most utilized \citep{Bearden1999} and seven to ten categories having some evidence as optimal \citep{Preston2000}.

Regardless of the number of options in a response scale, the use of discrete,  ordered categories can at times result in items with few or even no responses to some categories.
For example,  suppose individuals are asked to report the frequency in which a feeling arises, such as ``How often do you feel stable and secure in your life?'', and the individuals are provided with a four-point frequency response scale with labels 4 = ``Always, '' 3 = ``Often,'' 2 = ``Rarely, '' and 1 = ``Never.'' 
Generally speaking, one might reasonably expect some individuals from a large group to endorse each category. 
Yet, the frequency distribution of item response might depend on the individuals being sampled.
That is, a sample of individuals from developed countries may be less likely to endorse ``Never'' leading to a response distribution with sparse use of the ``Never'' response option.
Sparseness may be viewed as a characteristic of a particular sample, but the sparseness need not limit the utility of having more response options on the whole.
Ideally, we would aim for a method of analyzing this item (and the entire scale) that is consistent across samples, and a consistent, methodologically stable approach helps ensure inferences are comparable across samples, regardless of the nuanced responses of any one sample.

Alternative strategies for dealing with the sparse or missing response options may be used in practice. 
In the extreme case when an item response category was completely unused within a sample, a researcher may ignore the missing category and model the item as if the instrument was administered with one fewer response for the item. 
Ignoring an empty response category avoids the issue of estimating thresholds in item factor analysis without a finite maximum likelihood estimate. 
If the endorsement of a response option is low ($<2-4\%$),  the responses may also be collapsed with an adjacent response category, such as collapsing ``Strong Disagree'' and ``Disagree'' responses into a single category.
Unfortunately, either approach can negatively impact parameter estimation \citep{DiStefano2021, Savalei2011}.

An item, or items, with a sparse response category may not be a significant issue in single-group, single-time point analyses of response scales because the category may be dropped without much substantive information lost, but care should be taken to ensure the collapsed or dropped categories do not significantly alter the meaning of the interpretation.
Increased convergence and estimation are excellent, but collapsing categories may not be warranted when the focus is on comparing the measurement model across samples or time.
Multiple group analyses or longitudinal analyses of constructs may become quite difficult to interpret if the number of categories varies among groups.
For instance, testing the invariance of thresholds across groups would be impossible if the number of thresholds varies among groups for the same item(s). 

Testing equivalence of parameters across groups is necessary to ensure measurement is not biased by \textit{a priori} group differences\citep{Millsap2012, Meredith1993}. 
A necessary step in examining invariance within a factor analytic framework is examining the initial structure (i.e., configural invariance).
The permutation tests of configural invariance \citet{Jorgensen2018} may be misleading when a response category is not endorsed by chance on a permutation and the model fails to converge.
Such occurrences may limit evidence supporting the invariance of the configural structure.
A varying parameter space among groups is unlikely to be representative of the population,  which could allow the peculiarities of a messy sample to dictate which inferences are possible and could limit the replicability of scientific investigations.
To avoid varying parameter spaces among groups for any sample size,  developing an approach to account for response categories without endorsement is therefore necessary.

Ordered categorical data have been the focus of numerous methodological studies of different estimation methods for varying sample sizes, number of response categories, distribution of response categories, number of items, and so on. 
Due to the contributions of these investigations, recommendations for estimation methods are now available to researchers who need to estimate factor or structural models using data collected using Likert-type response scales. 
That is, unweighted least squares (ULS) and diagonally weighted least squares (DWLS) have been commonly recommended
\citep{Bandalos2014, Forero2009, DiStefano2014, Joreskog1988, Muthen1993, Muthen1997, Savalei2013, Shi2018}. 
Despite these advances, a growing research area involves using Bayesian methods to estimate latent variable models with categorical indicators. 
Adopting a Bayesian estimation approach allows researchers to incorporate their substantive expectations via prior distributions on the model parameters \citep{Gelman2013}. 
There are many potential benefits to this modeling approach, some of which we aim to present in this paper. 
From a scientific perspective, researchers may incorporate their expectations about the phenomena under investigation resulting in more robust and realistic conclusions about the results \citep{Stefan2022, Vanpaemel2012}.
A difficulty, though, is the use of Bayesian methods within the categorical SEM context due to the specification and estimation of thresholds for the underlying latent response variable.
The thresholds constrain the order of the response categories, which means items with more than two response categories have a dependence that must be accounted for in the estimation or disordered thresholds will occur.
 
The purpose of this work is the implementation of Bayesian latent variable models with categorical indicators when response categories are infrequently or not used by respondents. 
\citet{Betancourt2019} described using a Dirichlet prior for threshold parameters whereby all the thresholds for a single item simultaneously versus putting a prior on thresholds in an ordered sequence within an item. 
The Dirichlet prior reduces the uncertainty on the thresholds and effectively bounds the estimation leading to more efficient sampling. 
The technical details for this approach are described in more detail in the methods below. 
This approach contrasts with the default approach in commonly used software (e.g., blavaan and Mplus), where threshold priors are defined sequentially. 
Utilizing the more structured prior is expected to improve estimation performance, such as interval estimate coverage, when the indicators contain categories without any, or relatively few, responses. 

The remainder of this paper is outlined as follows.
First, we describe Bayesian categorical factor analysis in more detail by providing an introduction to our implementation of the latent variable models evaluated in this paper. 
Then, we describe the prior specification and the proposed induced-Dirichlet prior to help address infrequent endorsement.
The remainder of the work is broken up into three studies.
The first study illustrates the resulting posterior distributions across different prior specifications when data are simulated.
Study 2 is a Monte Carlo simulation study evaluating the posterior convergence, posterior coverage rates, and credible interval widths.
Study 3 illustrates the effects of sparse response patterns using the Gallup World Poll data on well-being through harmony in one's life.
We provide highlights from the results of the single simulated dataset, Monte Carlo study, and applied multi-group model example.
We conclude with a summary of our findings from our studies and our recommendations for future research and applications. 

\subsection*{Bayesian categorical factor analysis}

This paper focuses on using Bayesian latent variable models with categorical indicators. 
To begin, consider the measurement model for items $i=1,\ldots,I$ with response categories that possibly vary among items. Let $C_i$ represent the number of response categories for item $i$. 
In the measurement model for respondents, the latent variable $\bm\eta$ for $n=1,\ldots,N$ respondents can be denoted by
\begin{align*}
\mathbf{y}^\ast &= \bm\nu + \bm\Lambda\bm\eta + \bm\varepsilon\\
y_{ni} &= \begin{cases}
1,\ \text{if } y^\ast_{ni} < \tau_{i,1}\\
2,\ \text{if } \tau_{i,1} < y^\ast_{ni} < \tau_{i,2}\\
\vdots\\
C_i,\ \text{if } y^\ast_{ni} > \tau_{i,C_i-1}
\end{cases}
\end{align*}
In a Bayesian perspective, the model can be described distributionally by the latent variable $\bm\eta \sim \text{MVN}(\bm 0, \bm\Phi)$, $\bm\varepsilon \sim \text{MVN}(\bm 0, {\bm\Theta})$, $\bm\Lambda$ is a matrix of factor loadings with possibly many zero entries to define the different factors, and $\bm\Theta$ is the latent response residual variance-covariance matrix. Together, these result in the latent response distribution being denoted by $\bm{y}^\ast \sim \text{MVN}(\bm\nu + \bm\Lambda\bm\eta, {\bm\Theta})$. 
In this work, we followed the approach utilized in blavaan \citep{Merkle2021} by implementing the marginal likelihood of the latent variable model resulting in $\bm\eta$ not being directly sampled.
In the marginal likelihood, the above latent response distribution implies to $\bm{y}^\ast \sim \text{MVN}(\bm\nu, \bm\Lambda \bm\Phi\bm\Lambda^\prime + {\bm\Theta})$.
For single-group analyses, the likelihood is commonly further simplified by fixing $\bm\nu=\mathbf{0}$ for identification of the thresholds.
Implementing the marginal likelihood significantly simplifies the likelihood expression for more efficient sampling of the joint posterior. 

Sampling the latent response variable is nontrivial as the multivariate normal distribution has varying truncation points depending on the item, number of response categories, and observed response.
Implementing the sampling from the truncated multivariate normal distribution followed from the program written by \citet{Goodrich2017} based on the methods described by \citet{Geweke1994}.
The resulting implementation for ordinal indicators is a generalization of the multivariate probit model \citet{Goodrich2016} wrote for dichotomous data. 
The implementation of the truncated multivariate normal distribution is shown in our Online Supplement.
Next, the model prior structure is discussed in more detail as to how the Dirichlet prior specification works.

\subsection*{Model Priors}
\begin{figure}[!htp]
\centering
\includegraphics[width=0.75\textwidth]{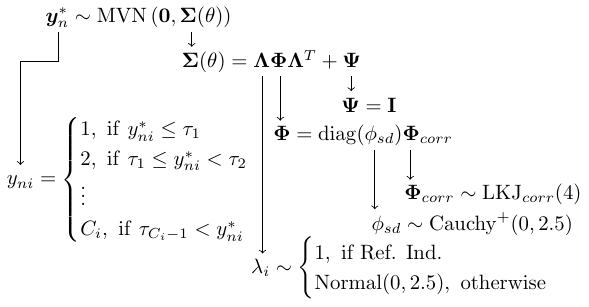}
\caption{Model-prior specification for a general item factor model.}
\label{fig:msd}
\vspace*{-0.25cm}
\begin{flushleft}
\textit{Note.} The prior specification for the threshold is intentionally left off of the above diagram.
\end{flushleft}
\end{figure}

The model prior choices below are based on the default settings in blavaan \citep{Merkle2016, Merkle2021}. 
The prior structure is shown in Figure \ref{fig:msd}.
The prior for the latent variable covariance matrix is decomposed into a Lewandowski-Kurowicka-Joe (LKJ) prior \citep{Lewandowski2009} for the correlations and independent half-Cauchy priors for the variance components.
The notable omission is the prior for the thresholds. 
The priors for the threshold parameters are described in the next section. 
	
\subsubsection*{Threshold Priors – Sequentially Defined}
Currently, a common default approach to setting the priors for thresholds is a set of normal priors on an unconstrained set of thresholds followed by a transformation to obtain an ordering to the thresholds \citep{Merkle2021}.
The approach in Mplus is to sequentially, or \textit{conditionally}, define the normally distributed threshold priors but using truncated distributions based on the previous threshold \citep[][, p. 11-16]{Asparouhov2010}. 
Implementing the truncated distribution approach of Mplus in Stan is complicated because Mplus uses different methods to sample the thresholds depending on the model.
The default prior in Mplus for thresholds is $\text{Normal}(0,10^{10})$ (parameterized using the variance) \citep[][, p. 34]{Asparouhov2010}, but how the truncation occurs is not clearly defined to the best of our understanding.

In this paper, we focused on the blavaan approach to obtaining ordered thresholds due to the implementation difficulties of the truncated approach in Stan.
For an item with $C$ response categories, the $C-1$ thresholds are specified as
\begin{align*}
\tau^\ast_c &\sim \text{Normal}(\mu_{\tau},\sigma_{\tau}),\ \text{for}\ c=1,\ldots,(C-1)\\
\tau_1 &= \tau^\ast_1\\
\tau_2 &= \tau^\ast_1 + \exp(\tau^\ast_2)\\
&\vdots\\
\tau_{C-1} &= \tau^\ast_1 + \exp(\tau^\ast_2) + \ldots +\exp(\tau^\ast_{C-1}).
\end{align*}
Sequentially defining the thresholds as a sum of exponentials of unconstrained parameters is a useful way to force an ordering to the threshold parameters to avoid boundary constraints when sampling parameters. 
The transformed parameters are then ordered. 
In blavaan, the default is $\tau^\ast_c \sim \text{Normal}(0,1.5)$, while in Mplus, the default for threshold parameters is essentially uniform over all real numbers, ``The default prior for all thresholds is Normal(0, $10^{10}$).'' \citep[][, p. 34]{Asparouhov2010}. 
The prior is not described as using a sum of exponentials to obtain an ordering, but instead, Mplus uses a truncation approach. 
How the truncated distributions are sampled is automatically selected based on the model, but the details on which is applied in a given model are unclear \citep[][, p. 11-16]{Asparouhov2010}. Therefore, due to the lack of clarity of the sampling of thresholds in Mplus, we chose to use the similar exponential sum blavaan uses but with the Mplus stated default prior.
We will use the large variance implied by the Mplus default as a comparison in our study but caution generalizing our results to Mplus directly due to the differences in sampling methods.

Placing informative priors on individual thresholds is not entirely obvious.
Consider, for instance,  an assessment on which items are scored with four ordered categories, and previous results are available to create informative priors for the thresholds.
The reported threshold point estimates for one of the items are $-$2.00, $-$0.25, and 1.75.
An informative prior is straightforward for the first thresholds, $\tau_1 \sim \text{Normal}(-2.00, 0.20).$
An informative prior for the second threshold is unfortunately not simply $\tau_2 \sim \text{Normal}(-0.25, 0.20)$, but is instead specified \textit{relative} to threshold 1. 
A more detailed demonstration of how the values can be obtained is provided in our Online Supplement.
A consequence of this prior structure is that not only are the expected values of the threshold priors ordered but the variances of the priors are also ordered such that
\begin{align*}
E[\tau_1] &< E[\tau_2] < E[\tau_3] < ...\\
Var[\tau_1] &< Var[\tau_2] < Var[\tau_3] < ...
\end{align*}
The equality of variances can only be achieved by fixing the variance of $Var[\tau^{\ast}]=0$, but this results in an improper prior.
The inflexibility of the aforementioned prior structure in allowing more precision to the prior of higher thresholds is limiting and may not align with realistic conditions.

\begin{figure}[!htp]
\centering
\includegraphics[width=0.99\textwidth]{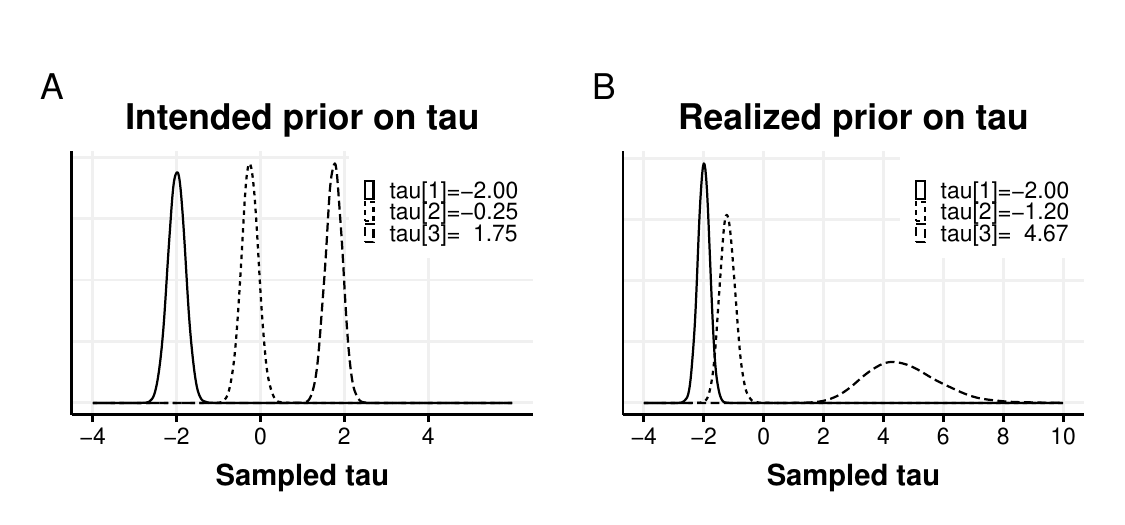}
\caption{Realized prior of ordered thresholds can be substantially different than intended}
\label{fig:prior-thresh-seq}
\begin{flushleft}
\textit{Note.} Parameters were simulated from priors with $N=10000$ plotted as a stacked histogram (binwidth=0.10). \textbf{A}. The intended informative prior on all three thresholds; \textbf{B}. Realized prior on the ordered thresholds. Intending to specify the informative priors $\tau_1\sim \text{Normal}(-2.00, 0.20)$, $\tau_2\sim \text{Normal}(-0.25, 0.20)$, and $\tau_3\sim \text{Normal}(1.75, 0.20)$ on the three ordered thresholds but failing to account for the sequential exponentials results in drastically different realized priors with $E[\tau_1] \approx -2.00$, $E[\tau_2] \approx -1.20$, and $E[\tau_3] \approx 4.50$.
\end{flushleft}
\end{figure}

Examples of realized prior distributions for the three ordered thresholds of a four-category item are shown in Figure \ref{fig:prior-thresh-seq}.
The figure illustrates the resulting prior on the three ordered thresholds after \textit{attempting} to place an informative prior on the thresholds.
The researcher intends to implement the informative priors $\tau_1\sim \text{Normal}(-2.00, 0.20)$, $\tau_2\sim \text{Normal}(-0.25, 0.20)$, and $\tau_3\sim \text{Normal}(1.75, 0.20)$ on the three thresholds.
However, due to how software implements the priors for thresholds, the realized prior on the thresholds is drastically different, especially for $\tau_3$ being significantly more diffuse and centered around approximately 4.50 instead of 1.75.
Failing to appropriately account for how the software transforms the specified prior can clearly result in an unintended realized prior that may not be close to what the researcher intended.
An alternative approach is to place a prior on the set of thresholds simultaneously, as discussed next, to avoid the confusion and difficulties inherent in the sequentially defined prior structure.

\subsubsection*{Threshold Priors – Induced-Dirichlet}

\citet{Betancourt2019} discussed an ``Induced-Dirichlet'' prior for ordered thresholds in ordered logistic regression. 
The prior was designed to produce robust inferences across a wider range of observed data characteristics, namely,  sparse responses. 
The prior is placed on all the thresholds simultaneously, resulting in a regularizing prior, which limits the likelihood of extreme values for the thresholds occurring in the posterior distributions. 
The nontrivial prior is as follows.

As described by \citet{Betancourt2019}, let $C_i$ be the number of response categories for item $i$ then an item will have $C_i-1$ threshold parameters ($\bm\tau=[\tau_1,\ldots,\tau_{(C_i-1)}]$) to estimate. 
The threshold parameters define a simplex, or a vector of probabilities, that sum to 1; that is, $\bm P=[P_1,P_2,\ldots,P_{(C_i)}]$ where $\sum_{c=1}^{C_i}P_c =1$. 
The proposed prior maps these probabilities onto the latent response variable, which in turn regularizes the thresholds indirectly by constraining the vector of probabilities.
In this paper, the mapping is accomplished using the normal CDF, $\Phi(.)$, and inverse normal CDF, $\Phi^{-1}(.)$. 
Therefore, the response probabilities are \textit{induced} by the mapping of $\tau\rightarrow P$, resulting in probabilities defined as
\begin{align*}
P_1 &= \Phi(\tau_1)\\
P_2 &= \Phi(\tau_2)-\Phi(\tau_1)\\
&\vdots\\
P_{C_i-1}&=\Phi(\tau_{C_i-1})-\Phi(\tau_{C_i-2})\\
P_{C_1}&=1-\Phi(\tau_{C_i-1}).
\end{align*}
The transformation from thresholds to probabilities requires an adjustment to the density function through the Jacobian matrix of partial derivatives. 
The reparameterization from thresholds to probabilities results in a change in the general density being sampled, which can, in turn, alter inferences if the adjustment to the calculated density is not accounted for \citep[see ][, for an excellent discussion of change of variables and Jacobian adjustments]{Carpenter2016}. 
The adjustment, in this case, follows taking the determinant of the Jacobian matrix. 
The partial derivatives come from taking the derivative of normal CDF with respect to the threshold parameters, which is the normal distribution PDF. Defining the partial derivative as $\phi(.)$, for ease of discussion, the Jacobian matrix reduces to
\begin{equation*}
J(\bm\tau_i) = \begin{pmatrix}
1 & \phi(\tau_1) & 0 & \cdots & 0\\
1 & -\phi(\tau_1) & \phi(\tau_2) & \cdots & 0\\
1 & 0 & -\phi(\tau_2) & \cdots & 0\\
\vdots & \vdots & \vdots & \ddots & \vdots\\
1 & 0 & 0 & \cdots & -\phi(\tau_{C_i-1})
\end{pmatrix}.
\end{equation*}
The probability density function combining the (1) Dirichlet prior for the probabilities given the thresholds and (2) the Jacobian adjustment for the change of variables from thresholds to probabilities results in the reduced form of
\begin{equation*}
Pr(\bm\tau \vert \bm\alpha) = \text{Dirichlet}(\bm p(\bm\tau)\vert \bm\alpha)\times\vert J(\bm\tau)\vert,
\end{equation*}
where $\bm\alpha=(\alpha_1,\alpha_2,\ldots,\alpha_{C_i})$ and $\vert J(\bm\tau)\vert$ represents the determinant of the Jacobian matrix above. 
For more details on the implementation of this prior in Stan, see our Online Supplement.
The prior on the vector of thresholds $\bm\tau$ is known as an induced-prior.

The induced-Dirichlet prior is characterized by the hyperparameter vector $\bm\alpha$. 
The vector $\bm\alpha$ represents the weight given to each response category. 
In the simplest case of the Dirichlet distribution with two categories, the distribution reduces to the univariate Beta distribution. 
The Beta distribution is commonly characterized using scalars alpha ($\alpha$) and beta ($\beta$), reflecting the relative probability of being one of these two categories. 
When the hyperparameters $\alpha=\beta=1$, the Beta distribution is known as a flat or uniform prior over probabilities.
A common interpretation of the Beta prior hyperparameters in terms of sample size is an additional $\alpha$ responses to category 1 and $\beta$ additional responses to category 2. 
The induced-Dirichlet's prior hyperparameter vector $\bm\alpha$ extends this interpretation to more than two categories, reflecting the number of additional responses to each response category.

\begin{figure}[!htp]
\centering
\includegraphics[width=0.98\textwidth]{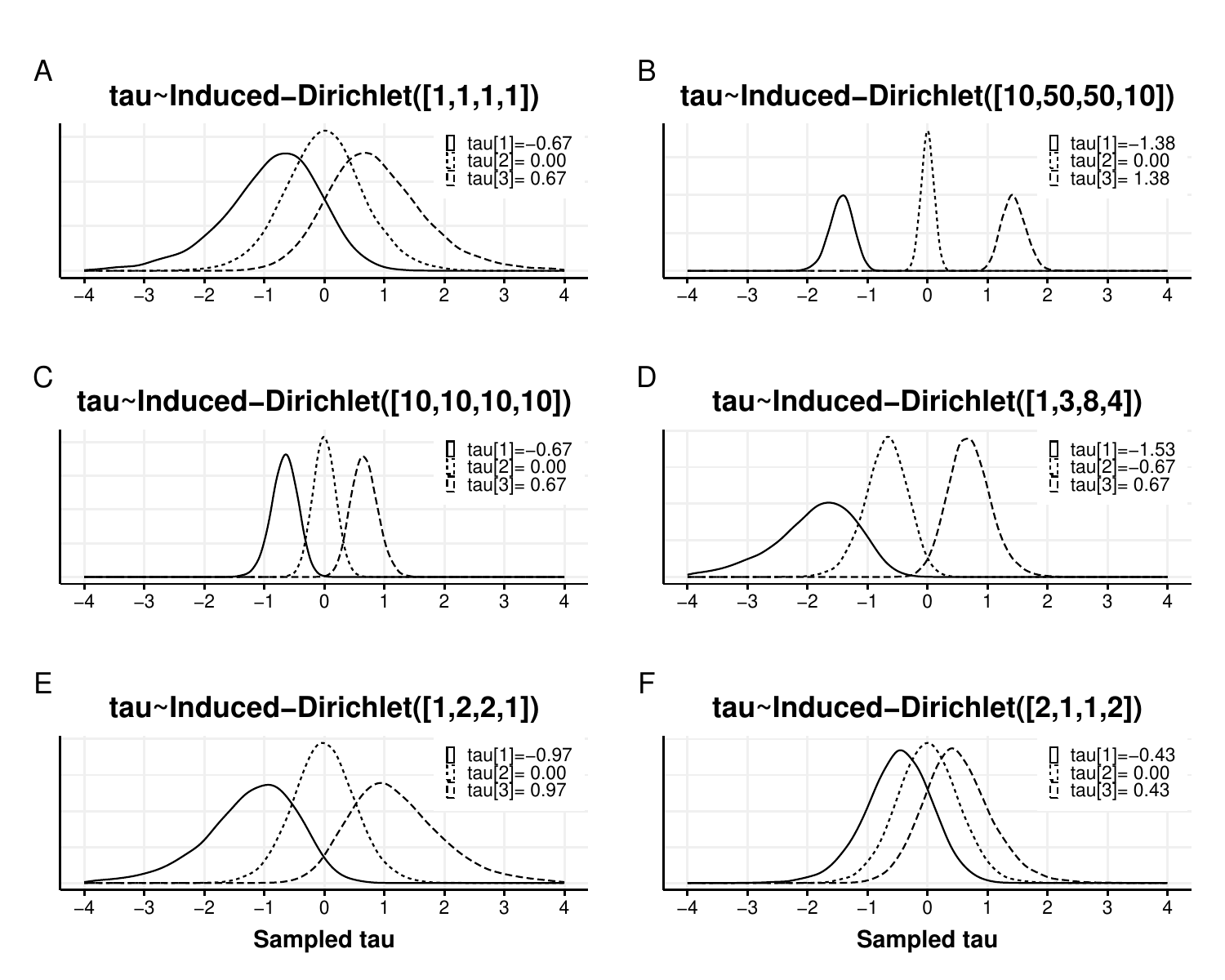}
\caption{Realized prior of ordered thresholds using a joint induced-Dirichlet prior}
\label{fig:prior-thresh-joint}
\begin{flushleft}
\textit{Note.} Priors were plotted as a stacked histogram (binwidth=0.05). \textbf{A}. $\bm\alpha = (1,1,1,1)$ is our initial suggestion for a relatively uninformative specification of the induced-Dirichlet prior; \textbf{B}. $\bm\alpha = (10,10,10,10)$ illustrates that more precision can be induced to each threshold by increasing the weight uniformly; \textbf{C}. $\bm\alpha = (10,50,50,10)$ illustrates how greater precision can be induced to the middle threshold only by increasing the middle weights but not the extremes; \textbf{D}. $\bm\alpha = (1,3,8,4)$ illustrates a relatively informative (relative to A) prior with more weight given to the upper response categories; \textbf{E}. $\bm\alpha = (1,2,2,1)$ illustrates a relatively uninformative prior for thresholds while still giving more weight to central response categories; and \textbf{F.} $\bm\alpha = (2,1,1,2)$ illustrates a relatively uninformative prior for thresholds that gives more weight to the extreme responses with the intent of a relatively weak regularizing effect.
\end{flushleft}
\end{figure}

Reconceptualizing the prior regularizes the estimation of the thresholds due to the constraints placed on the underlying probabilities through the Dirichlet distribution.
Modifying the values in the prior hyperparameters $\bm\alpha$ provides differing degrees of regularization.
For example, with four response categories, a uniform prior over the response probabilities is obtained using $\bm\alpha=(1,1,1,1)$ and provides equal weight to each response category.
Modifying the vector to $\bm\alpha = (1,2,2,1)$ induces a greater weight to the middle response categories than either extreme, thus increasing the distance between thresholds.
Modifying the vector to $\bm\alpha = (2,1,1,2)$ induces a greater weight to the extreme response categories, thus decreasing the distance between thresholds.
The more \textit{prior} weight given to the extreme response categories, the greater the regularizing effect the prior has on the posterior distribution of the extreme thresholds.
Examples of alternative priors hyperparameters $\bm\alpha$ with the induced thresholds are illustrated in Figure \ref{fig:prior-thresh-joint}.

Choosing among possible prior specifications can be challenging without an inferential interpretation of the prior hyperparameters $\bm\alpha$.
One way of interpreting the values of $\bm\alpha$ is based on the idea that each value represents a ``pseudo''-observation.
Specifying $\bm\alpha=(1,1,1,1)$ is then saying that in a sample of 4 people, we would expect each person to select a different category.
However, specifying $\bm\alpha=(10,10,10,10)$ similarly implies equal weight to each category, but now we are saying the equal weight is based on a sample of 40 ``pseudo''-observations, thus the prior is given more weight.
This interpretation of pseudo-observations becomes useful when constructing informative priors.
An informative prior for thresholds can be described in terms of sample size and the number of responses to each category given that same size.
For instance, the induced threshold distribution in Figure \ref{fig:prior-thresh-joint}C conveys the informative prior on a four-category item with 120 additional observations added to the estimation where 10 observations endorsed category 1, 50 observations endorsed category 2, 50 observations endorsed category 3, and 10 observations endorsed category 4.
However, a relatively less informative prior is shown in Figure \ref{fig:prior-thresh-joint}D.
Panel D is interpreted as in a sample of 16 additional respondents, we would expect 1 observation to endorse category 1, 3 observations to endorse category 2, 8 observations to endorse category 3, and 4 observations to endorse category 4.
Interpreting the Dirichlet prior in terms of ``pseudo''-observations is common in latent class models \citep{Galindo2006, Depaoli2022} and generalized linear models \citep{Schafer1997}.
Validating this interpretation is out of the scope of this work as our focus is on the estimation and implementation, but future work will aim to provide evidence of the prior equivalent sample size interpretation.

An important difference between this joint induced-Dirichlet prior and the sequential sum of exponential prior discussed previously is the lack of strict ordering of the variance associated with each threshold distribution. The joint prior allows each threshold to have different certainty, with middle thresholds typically being more precise than extreme thresholds. This conceptually aligns with the fact small changes in central thresholds imply larger differences in response probabilities than small changes in extreme thresholds.

\subsection*{Research Questions}

The potential specifications of priors for thresholds in item factor analysis are examined in this paper.
Study 1 examines a single simulated dataset with known population parameters and one item per factor with a response category without endorsement. 
We illustrate how different prior specifications for thresholds (sequentially defined normal priors or a joint Induced-Dirichlet prior) influence how interpretable the resulting posterior distributions are for thresholds.
We expect that using the induced-Dirichlet prior will result in an interpretable posterior even when a category is not endorsed (as simulated). 
A large variance for the sequentially defined priors is expected to yield a very diffuse posterior for the threshold without endorsement, as no information pulls in the extreme values.

Study 2 investigates under what conditions the different prior structures perform optimally in terms of posterior convergence, credible interval coverage, and widths of credible intervals (i.e., efficiency). 
Again, the induced-Dirichlet prior is expected to perform well under all conditions, even those with messy data characteristics (e.g., missing cells).
In the conditions when respondents sufficiently endorse all categories, we expect each prior specification to perform well.

In study 3, we demonstrate the effects of using different prior structures in a multiple-group item factor analysis using data from the Gallup World Poll. 
We demonstrate the effects of prior specification on the resulting threshold posterior distributions for each group.

\section*{Study 1: Illustrative Example}
\subsection*{Methods} 
\begin{figure}[!htp]
\centering
\includegraphics[width=0.5\textwidth]{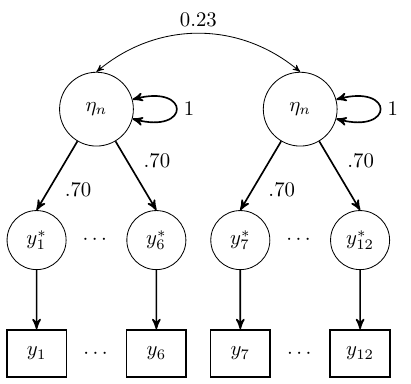}
\caption{Path diagram representation for a two-factor model with population values for factor loadings}
\label{fig:path-model}
\vspace*{-0.25cm}
\begin{flushleft}
\textit{Note.} The above diagram illustrates the population model used for data generation. The factor loadings reported are standardized, where 0.70 was found to be the average reported standardized loadings in a recent literature review \citep{Padgett2022}. Residual variances of the latent response distributions were fixed at $\theta=1-0.7^2=0.51$. Additional details on data simulation are available in our online supplement.
\end{flushleft}
\end{figure}

A single dataset based on a simple two-factor model was generated with a relatively small sample size of 150; the lower end of sample sizes was selected from a review of confirmatory factor analysis (CFA) applications \citep{Padgett2022}.
The population parameter values used to simulate these data were factor variances equal to one, a factor correlation of 0.23, latent response residual variances equal to one, and factor loadings equal to one.
These parameters result in standardized factor loadings of approximately 0.70. 
One item per factor was generated to have a ``sparse'' response scale where one category was not endorsed. 
The sparse items were created by using thresholds [$-$15.0, -1.25, -0.25], leading to category response probabilities of [0, 0.19, 0.24, 0.57].
All non-sparse items were generated using thresholds [$-$2.32, -1.25, -0.25], leading to category response probabilities of [0.05, 0.14, 0.24, 0.57].
All items were supposed to have four response categories.
The structure of the model is shown in Figure \ref{fig:path-model}.

The model is estimated using the prior structure defined in the Model Prior section and using five alternative priors for the thresholds: (1) sequential log-normal with a standard deviation of 1.5, (2) sequential log-normal with a standard deviation of 10000, (3) joint Induced-Dirichlet with  $\alpha=[1,1,1,1]$, (4) joint Induced-Dirichlet with $\alpha=[7,21,36,86]$, and (5) joint Induced-Dirichlet with $\alpha=[37,37,37,37]$.
The last two Induced-Dirichlet priors were highly informative, where the hyperparameters sum to sample size. The first is an ``accurate'' informative prior, and the last is a ``misspecified'' informative prior, giving equal weight to each category.
Both priors are more informative than would generally be used in a model, but the extreme weight illustrates the regularizing effect of the Dirichlet prior.
The parameter estimates of each model were evaluated to investigate differences among the posterior distributions of thresholds using the three different prior structures.
The R script and Stan model files for replicating this study are available on our accompanying OSF repository (\url{https://osf.io/kpvwb/}).

\subsection*{Results}

The resulting credible intervals (CI) of the posterior distributions of the thresholds of items one and eight across the three prior specifications are shown in Table \ref{tb:sim1-res}. 
The major differences in the posterior distribution for the first threshold are of particular attention.  
An uninformative prior resulted in the posterior of the first threshold being quite wide, given that the observed data did not update the prior meaningfully. 
On the other hand, the difference between the resulting CI for the Induced-Dirichlet prior compared to the Normal(0,1.5) prior appears negligible. 
Despite the similarity in the posterior distributions for the two relatively informative priors, differences may exist in sampling efficiency, which the Monte Carlo simulation study will help to tease apart. 

\begin{table}[!htp]
\centering
\small
\begin{threeparttable}
\caption{Posterior distribution of sparse items illustrates the regularizing effect of informative induced-Dirichlet priors.}
\label{tb:sim1-res}
\begin{tabular}{l rrr}
\toprule
Prior & Threshold 1 ($\tau_1$) & Threshold 2 ($\tau_2$) & Threshold 3 ($\tau_3$) \\ \midrule
\multicolumn{4}{c}{Item 1 Thresholds - All Categories Endorsed}\\
Population Value & $-$2.32 & $-$1.25 & $-$0.25 \\
PML Est. & $-$2.61 ($-$3.27,$-$1.95) & $-$1.31 ($-$1.66,$-$0.96) & $-$0.25 ($-$0.53,0.03)\\
Large Variance & $-$2.36 ($-$2.90,$-$1.77) & $-$1.20 ($-$1.50,$-$0.89) & $-$0.23 ($-$0.49,0.01) \\
Small Variance & $-$2.32 ($-$2.92,$-$1.82) & $-$1.18 ($-$1.51,$-$0.88) & $-$0.20 ($-$0.45,0.07)\\
$\text{ID}\left([1,1,1,1]\right)$ & $-$2.25 ($-$2.81,$-$1.77) & $-$1.09 ($-$1.41, $-$0.79) & $-$0.11 ($-$0.36, 0.12)\\
$\text{ID}\left([7,21,36,86]\right)$ &  $-$2.07 ($-$2.44,$-$1.72) & $-$0.99 ($-$1.18, $-$0.81) & $-$0.17 ($-$0.31, $-$0.02)\\
$\text{ID}\left([37,37,37,37]\right)$ & $-$0.86 ($-$1.07,$-$0.66) & $-$0.10 ($-$0.28, 0.06) & 0.68 (0.49, 0.88)\\
~~\\
\multicolumn{4}{c}{Item 2 Thresholds - Category 1 Not Endorsed}\\
Population Value & $-$15.0 & $-$1.25 & $-$0.25 \\
PML Est. & ${}^{\ast}$ & $-$1.21 ($-$1.54,$-$0.87) & $-$0.23 ($-$0.47,0.02) \\
Large Variance & $-$75.9 ($-$402.5,$-$3.58) & $-$1.35 ($-$1.71,$-$1.02) & $-$0.23 ($-$0.54,0.02) \\
Small Variance & $-$3.34 ($-$4.65,$-$2.43) & $-$1.23 ($-$1.58,$-$0.91) & $-$0.21 ($-$0.48,0.03)\\
$\text{ID}\left([1,1,1,1]\right)$ & $-$3.47 ($-$5.27,$-$2.42) & $-$1.14 ($-$1.47,$-$0.85) & $-$0.13 ($-$0.38, 0.12)\\
$\text{ID}\left([7,21,36,86]\right)$ & $-$2.37 ($-$2.88,$-$1.94) & $-$1.02 ($-$1.21, $-$0.84) & $-$0.18 ($-$0.32, $-$0.04)\\
$\text{ID}\left([37,37,37,37]\right)$ & $-$0.93 ($-$1.15,$-$0.73) & $-$0.12 ($-$0.29, 0.04) & 0.67 (0.49, 0.85)\\
  \bottomrule
\end{tabular}
 \vspace*{1mm}
 	\begin{tablenotes}[para, flushleft]
    {\small
        \textit{Note.} $N=150$; ${}^{\ast}$Parameter not estimated by lavaan due to no responses observed in category 1; models were estimated using the THETA parameterization where the model implied latent response variable variance is 2; PML Est. = Pairwise Maximum Likelihood Point Estimate from lavaan \citep{lavaan}; Large variance=$\text{Normal}(0, 10^5)$; Small Variance = $\text{Normal}(0, 1.5)$ default prior in blavaan \citep{Merkle2021}; $\text{ID}\left([1,1,1,1]\right)$ corresponds to a diffuse Induced-Dirichlet prior; $\text{ID}\left([7,21,36,86]\right)$ corresponds to an informative Induced-Dirichlet prior with response probabilities approximately [0.05, 0.15, 0.25, 0.55];  and $\text{ID}\left([37,37,37,37]\right)$ corresponds to an informative Induced-Dirichlet prior with equal weight given to each category. 
    }
 	\end{tablenotes}
 \end{threeparttable}
\end{table}

%
%
%

\section*{Study 2: Monte Carlo Simulation}
\subsection*{Methods}
\begin{figure}
\centering
\includegraphics[width=0.95\textwidth]{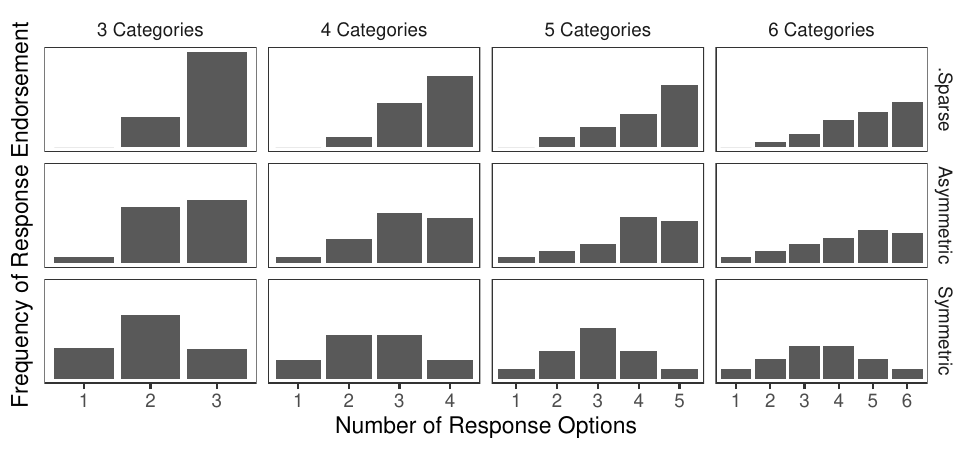}
\caption{Category response distributions simulated in Monte Carlo study}
\label{fig:cat-dist}
\end{figure}

Secondly, a Monte Carlo simulation study was conducted to evaluate the estimation performance of the proposed prior structure. 
The design of this Monte Carlo simulation varied the distribution of the indicators (i.e., symmetry of category endorsement), the number of response categories (i.e., 3, 4, 5, 6), and sample size (i.e., 150, 500). 
The set of response distribution conditions is shown in Figure \ref{fig:cat-dist}. 
For the sparse category condition, three additional conditions were added: 1) all items contain a category without endorsement, 2) half of all items are simulated with a category without endorsement, and 3) each factor contains one item with a category without endorsement. 
The non-sparse items in these conditions were simulated using the asymmetric distribution condition.
The number of response categories and sample size conditions were determined based on a recent literature review by \citet{Padgett2022}, which indicated that these values are commonly observed in applied research in psychological assessment. 

This study focused on the priors to the sequential normal priors (small and large variance) and the relatively diffuse Induced-Dirichlet prior with unit hyperparameter. The focused priors provide a basis for further investigations of more nuanced specifications of the Dirichlet prior. Therefore, the Induced-Dirichlet prior specification was kept simple in all conditions with unit hyperparameters regardless of the number of categories. Leading to the interpretation that the prior adds 1 response to each category.

The posterior distributions were sampled using Stan with four chains and 2000 iterations per chain. 
The first 1000 iterations were discarded as warm-up. 
The results of the simulation focus on the coverage rates and interval widths of 95\% credible intervals for all parameters and posterior convergence measured by the effective sample size and $\hat{R}$. 
Coverage rates were computed by summarizing the frequency with which the true value lay within the central 95\% quantiles of the posterior distributions.
The interval widths were computed using the difference between the 97.5-and 2.5 percentiles of the posterior distributions.
Evaluating coverage rates and interval widths informed how well the posterior distributions captured the population parameters. 
Preferred are credible intervals with well-calibrated coverage rates ($\approx$95\%) and where the credible interval widths are narrower. 
The results of the Monte Carlo simulation were summarized by averaging over replications within each condition to obtain the average coverage rates and the average credible interval widths across conditions.
Additional results describing convergence, coverage rates, and credible interval widths are reported in our Online Supplement with complete stan files in our online repository (\url{https://osf.io/kpvwb/}).

To obtain an empty response category in our simulation, we fixed the first threshold of the appropriate items to $-$15, which induced a probability of anyone selecting the first category to essentially zero. 
In our summary of results of coverage or credible interval widths across replications and conditions, we refer to a special set of results where the first threshold was fixed, and we have labeled such conditions as ``Empty Category $1^{st}$ Threshold.''
The results in these conditions refer to estimating a parameter that wouldn't exist under maximum likelihood estimation, and the simulating value is $-$15.
However, using a regularizing prior, we aim to constrain the estimated uninformed threshold to a \textit{reasonable} value relative to other thresholds for that item, where \textit{reasonable} would be an estimated threshold within the range of commonly encountered values for thresholds in item factor models, such as in the range of (-4,-2) under a standardized solution. When estimating the coverage rates, we used the unstandardized samples from the posterior distribution to evaluate where the population thresholds were contained within the corresponding 95\% credible intervals.

\subsection*{Results}

\subsubsection*{Posterior convergence}
Posterior convergence was assessed using the effective sample size and Gelman-Rubin-Brooks convergence criteria, $\hat{R}$.
Averaging over all parameters and conditions,  the joint induced-Dirichlet prior resulted in an average $\hat{R}=1.007$ with 98.7\% of estimates below 1.1 and an average effective sample size of 2479. The small variance prior resulted in an average $\hat{R}=1.004$ with 99.4\% of estimates below 1.1 and an average effective sample size of 2470.  The large variance prior resulted in an average $\hat{R}=1.148$ with 64.2\% of estimates below 1.1 and an average effective sample size of 864.
Under these conditions, the large variance prior distribution resulted in significantly poorer summary estimates of convergence than the smaller variance prior.
The posterior distributions converged on average except under the large variance prior.

Of special interest in this study are the posterior distributions for the thresholds.
Summaries of the threshold parameter convergence criteria across prior distributions and response distributions are given in Table \ref{tb:converg-thresh}.
An asymmetric or symmetric response distribution resulted in adequate convergence regardless of prior choice, but if even one item is sparsely endorsed the small variance or joint prior converged much better than the large variance prior.
In summary of convergence, the difference between the joint induced-Dirichlet and small variance successive sum prior was negligible, but the large variance successive sum prior converged poorly.

\begin{table}[!htp]
\centering
\small
\begin{threeparttable}
\caption{Convergence of threshold posterior depended on the distribution of observed indicators and prior specification}
\label{tb:converg-thresh}
\begin{tabular}{llccr}
  \toprule
Distribution\\ (\# Sparse Items) & Prior & Avg. $\hat{R}$ & \% $\hat{R}<1.1$ & Avg. ESS\\ 
  \midrule
  Symmetric (0) & Joint & 1.01 & 0.99 & 2552.7 \\ 
    & Small Variance & 1.01 & 0.99 & 2492.4 \\ 
    & Large Variance & 1.01 & 0.99 & 2457.1 \\
  Asymmetric (0) & Joint & 1.00 & 1.00 & 2296.9 \\ 
    & Small Variance & 1.00 & 1.00 & 2295.3 \\ 
    & Large Variance & 1.00 & 1.00 & 2191.4 \\ 
  Sparse (2) & Joint & 1.01 & 0.98 & 2569.1 \\ 
    & Small Variance & 1.00 & 1.00 & 2483.2 \\ 
    & Large Variance & 1.19 & 0.48 & 133.8 \\ 
  Sparse (6) & Joint & 1.01 & 0.98 & 2513.3 \\ 
    & Small Variance & 1.00 & 0.99 & 2444.9 \\ 
    & Large Variance & 1.21 & 0.47 & 125.9 \\ 
  Sparse (12) & Joint & 1.01 & 0.98 & 2434.2 \\ 
    & Small Variance & 1.00 & 0.99 & 2659.9 \\ 
    & Large Variance & 1.26 & 0.46 & 155.9 \\ 
   \bottomrule
\end{tabular}
 \vspace*{1mm}
 	\begin{tablenotes}[para, flushleft]
    {\small
        \textit{Note.} Joint = joint induced-Dirichlet prior; Small Variance = $\text{Normal}(0, 1.5)$; Large variance=$\text{Normal}(0, 10^5)$; Avg. ESS = average effective sample size.
    }
 	\end{tablenotes}
 \end{threeparttable}
\end{table}

\subsubsection*{Posterior coverage}

The coverage rates of the posterior credible intervals is considered next.
Averaging over all conditions and parameters except the fixed threshold parameters,  coverage rates below the nominal level 0.95 for each prior structure were: 0.757 for the joint prior, 0.789 for the small variance prior, and 0.784 for the large variance prior.
Across conditions and parameters, the coverage rates for the prior specifications varied from 0.13 to 0.99 for the joint prior, 0.14 to 0.99 for the small variance prior, and 0.15 to 0.99 for the large variance prior.
The high variability in coverage rates among conditions and parameters indicates that coverage is potentially problematic regardless of prior specification when sample sizes are 500 or less.

The coverage rates varied substantially among conditions, and we identified substantial differences in coverage among the different response distributions (Table \ref{tb:coverage}).
We found negligible differences in the coverage rates and average credible interval widths across the prior specifications within the symmetric and asymmetric response distribution conditions for all parameters.
However, excluding the fixed thresholds, the CI coverage was significantly worse in the asymmetric and sparse response distributions compared to the symmetric response distribution conditions.
This pattern was consistent across all number of response categories conditions.
Within the sparse response distribution conditions, we found the number of items that contained an empty response category had a major impact on the coverage rates.

\subsubsection*{Posterior CI widths}

The average widths are reported in Table \ref{tb:coverage}.
The average CI widths were consistent across priors specifications within the symmetric and asymmetric response distribution conditions for each parameter.
We found substantial differences in the CI widths across prior specifications in the sparse response distribution conditions.
The magnitude of the difference in CI width depended on the parameter and number of sparse items, which we focus on next.

Factor loading CI widths were approximately the same in the joint and small variance threshold prior specifications.
The CI widths for factor loadings were noticeably narrower in the large variance threshold prior specifications regardless of the number of sparse items.
However, the coverage rate for factor loadings also tended to be slightly lower for the large variance specifications compared to the other prior specifications which results in a trade-off between the coverage and interval widths depending on the prior specification for threshold parameters.

The factor variance patterns had a similar pattern of CI widths and coverage compared to the factor loadings.
The difference between these two parameters was that the difference in interval widths was less drastic.
The factor covariances tended to have similar CI widths but different coverage rates across Sparse response distribution conditions.

The threshold parameters highlight the severe differences among the CI widths between prior specifications.
For threshold parameters in items where all categories were endorsed, the CI widths were similar across prior specifications even when some of the other items within the factor were spare.
When all twelve items were generated with a sparse response distribution, the large variance prior had wider CIs than the small variance prior. 
However, for the threshold separating the empty response category and the next category, the large variance prior had impractical CI widths regardless of the number of sparse items.
In the conditions with only two sparse items, the average CI width of the threshold posteriors was over 56 units.
The widths under the joint induced-Dirichlet or small variance prior were more reasonable between 2-3. 
This result highlights how even a marginally informative prior with a reasonable variance can significantly impact the posterior distribution when an item contains a category without endorsement.

\newpage
\begin{landscape}
\begin{table}[!htp]
\centering
\begin{threeparttable}
\caption{Posterior coverage and CI width depended on prior and response distribution}
\label{tb:coverage}
\begin{tabular}{llcccccc}
  \toprule
  & Distribution & \multicolumn{3}{c}{Coverage Rate (\%)} & \multicolumn{3}{c}{Avg CI Width}\\ \cmidrule(lr){3-5} \cmidrule(lr){6-8}
 Parameter & (\# Sparse Items) & Joint & Small Variance & Large Variance & Joint & Small Variance & Large Variance \\ 
  \midrule
  Loadings & Symmetric (0) & 93.7 & 93.4 & 92.9 & 0.68 & 0.69 & 0.69 \\ 
   & Asymmetric (0) & 94.2 & 94.1 & 93.5 & 0.75 & 0.75 & 0.76 \\ 
   & Sparse (2) & 91.9 & 91.8 & 93.1 & 0.86 & 0.86 & 0.31 \\ 
   & Sparse (6) & 93.5 & 93.3 & 92.9 & 0.83 & 0.82 & 0.30 \\ 
   & Sparse (12) & 94.5 & 94.3 & 91.9 & 0.93 & 0.93 & 0.29 \\ 
  Factor Variances & Symmetric (0)  & 92.2 & 92.0 & 92.5 & 0.91 & 0.93 & 0.94 \\ 
   & Asymmetric (0) & 89.6 & 90.1 & 90.1 & 0.97 & 0.97 & 1.00 \\ 
   & Sparse (2)  & 86.9 & 87.5 & 85.0 & 0.95 & 0.95 & 0.81 \\ 
   & Sparse (6)  & 87.9 & 88.3 & 77.8 & 0.99 & 0.97 & 0.75 \\\
   & Sparse (12)  & 86.0 & 85.4 & 76.5 & 1.01 & 1.00 & 0.72 \\ 
  Factor Covariance & Symmetric (0)  & 90.9 & 90.5 & 90.6 & 0.29 & 0.30 & 0.30 \\ 
   & Asymmetric (0)& 89.5 & 90.3 & 89.8 & 0.30 & 0.30 & 0.31 \\ 
   & Sparse (2)  & 84.0 & 81.4 & 78.2 & 0.28 & 0.28 & 0.27 \\ 
   & Sparse (6)  & 88.8 & 88.4 & 80.0 & 0.30 & 0.29 & 0.28 \\ 
   & Sparse (12)  & 85.4 & 83.5 & 76.3 & 0.31 & 0.30 & 0.27 \\ 
  Thresholds & Symmetric (0) & 94.8 & 94.1 & 94.4 & 0.55 & 0.57 & 0.57 \\ 
   & Asymmetric (0) & 90.3 & 94.1 & 94.6 & 0.58 & 0.59 & 0.62 \\ 
   & Sparse (2)  & 88.9 & 93.8 & 92.6 & 0.58 & 0.59 & 0.52 \\  
   & Sparse (6)  & 87.8 & 93.5 & 92.5 & 0.58 & 0.59 & 0.52 \\ 
   & Sparse (12)  & 19.6 & 14.8 & 16.2 & 1.17 & 1.00 & 1.20 \\ 
  Empty Category & Sparse (2) & $^{\ast}$ & $^{\ast}$ & $^{\ast}$ & 3.21 & 2.13 & 56.76 \\ 
  $1^{st}$ Threshold$^{\ast}$ & Sparse (6)  & $^{\ast}$ & $^{\ast}$ & $^{\ast}$ & 2.89 & 2.15 & 44.51 \\
   & Sparse (12)  & $^{\ast}$ & $^{\ast}$ & $^{\ast}$ & 2.89 & 2.17 & 11.35 \\ 
   \bottomrule
\end{tabular}
 \vspace*{1mm}
 	\begin{tablenotes}[para, flushleft]
    {\small
        \textit{Note.} $^{\ast}$Threshold was set to $-15$ for simulating a ``sparse''/empty response category, and credible intervals for this threshold should NOT contain the simulating parameter. Joint = joint induced-Dirichlet prior; Small Variance = $\text{Normal}(0, 1.5)$; Large variance=$\text{Normal}(0, 10^5)$.
    }
 	\end{tablenotes}
 \end{threeparttable}
\end{table}

\end{landscape}
\newpage

\subsubsection*{Reference indicator effect}

In our initial plan of conditions in the simulation study, we set the sparse response distribution conditions to always use an item with an empty response category as the reference indicator of the factor. 
After we investigated the coverage and direction of mis-coverage (positive vs. negative bias) for parameters, we observed that the factor variance parameters were consistently underestimated in the sparse response distribution conditions. 
Our hypothesis was that using a variable with a more restricted range may cause underestimation.
To counter this potential issue, we reran conditions with a sparse response category for half the items and flipped which item was used as the reference indicator to a non-sparse item.
\begin{table}[!htb]
\centering
\small
\begin{threeparttable}
\caption{Using a reference indicator with an empty response category results in poorer coverage}
\label{tb:ind-variable}
\begin{tabular}{llccc}
  \toprule
  Parameter & Reference Indicator & Joint & Small Variance & Large Variance \\ 
  \midrule
  \multicolumn{5}{c}{\textit{Coverage Rate} (\%)}\\
  Loading & Sparse & 92.3 & 92.0 & 93.1 \\ 
   & Non-sparse & 94.6 & 94.7 & 92.8 \\ 
  Factor Variance & Sparse & 85.6 & 85.9 & 80.8 \\ 
   & Non-sparse & 90.3 & 90.7 & 75.0 \\ 
  Factor Covariance & Sparse & 87.8 & 86.9 & 78.4 \\ 
   & Non-sparse & 89.8 & 89.9 & 81.4 \\ 
  Thresholds & Sparse & 86.2 & 92.9 & 92.3 \\ 
   & Non-sparse & 89.2 & 94.0 & 92.7 \\ 
  Empty Category $1^{st}$ Threshold & Sparse & $^{\ast}$ & $^{\ast}$ & $^{\ast}$ \\ 
   & Non-sparse & $^{\ast}$ & $^{\ast}$ & $^{\ast}$ \\  \midrule
  \multicolumn{5}{c}{\textit{Avg. CI Width}}\\
  Loading & Sparse & 0.89 & 0.89 & 0.29 \\ 
   & Non-sparse & 0.76 & 0.75 & 0.31 \\ 
  Factor Variance & Sparse & 0.96 & 0.95 & 0.75 \\ 
   & Non-sparse & 0.98 & 0.98 & 0.75 \\ 
  Factor Covariance & Sparse & 0.29 & 0.28 & 0.28 \\ 
   & Non-sparse & 0.31 & 0.31 & 0.28 \\ 
  Thresholds & Sparse & 0.58 & 0.59 & 0.53 \\ 
   & Non-sparse & 0.58 & 0.59 & 0.52 \\
  Empty Category $1^{st}$ Threshold & Sparse & 2.89 & 2.15 & 98.45\\ 
   & Non-sparse &  2.89 & 2.15 & 28.13 \\ 
   \bottomrule
\end{tabular}
 \vspace*{1mm}
 	\begin{tablenotes}[para, flushleft]
    {\small
        \textit{Note.} $^{\ast}$Conditions (labeled ``Empty Category $1^{st}$ Threshold'') where the first threshold was set to $-15$ for simulating a ``sparse''/empty response category and credible intervals for this threshold should NOT contain the simulating parameter. Joint = joint induced-Dirichlet prior; Small Variance = $\text{Normal}(0, 1.5)$; Large variance=$\text{Normal}(0, 10^5)$.
    }
 	\end{tablenotes}
 \end{threeparttable}
\end{table}
The coverage rates and CI widths comparing the sparse and non-sparse are shown in Table \ref{tb:ind-variable}.
We found that using a reference indicator with an empty response category consistently resulted in poorer coverage rates for most parameters.
The CI widths for factor loadings tended to be narrower when a non-sparse item was used, but the CI widths did not substantially differ for the other parameters.
The sole exception was the CI width for the threshold separating the unendorsed category and the next category.

\section*{Study 3: Gallup World Poll}
\subsection*{Methods}

In this applied example, we demonstrate the effects of sparse response categories on how threshold posteriors may update depending on group membership in Bayesian item factor analysis. 
We used a subset of data from the 2022 Gallup World Poll, looking in particular at a module of eight items centered on balance and harmony \citep[see ][, for details on the development of the module]{Lomas2022}, which are integral to well-being and flourishing more broadly \citep{Lomas2021}. 
The survey usually takes 15-20 minutes to complete, featuring around 60 – 80 items (with the number varying among respondents based on screener questions, filters and skip patterns), and involves nationally representative, probability-based samples among the adult populations, aged 15 and older, of around 1,000 people per country (with some larger countries having between 2,000-3000, and some smaller countries having only 500). 
This sample size is intended to allow, after accounting for the survey weights, a maximum confidence interval of approximately four percentage points, providing enough power ($\beta=0.80, \alpha=.05$) to detect a group difference of approximately nine percentage points. 
Our analysis specifically focused on differences in balance and harmony between the United States sample (N = 1006) and the Norway sample (N = 1002).
 
Most items in the poll have binary yes/no response options, although psychometric surveys tend to have Likert-type scales, which allow more nuanced responding patterns and statistical analysis. 
However, Gallup has found through experience that yes/no items work better in the context of their international surveys for numerous reasons, including being easier for some participants to understand and being more readily standardized across cultures. 
To the latter point, for instance, cross-cultural variation has been observed in the way people from different cultures respond to Likert scales. 
For example, people in more individualistic societies seem to show a preference for response options at the extremes, potentially because it allows them to stand out personally, whereas people in more collectivistic societies hew more towards the center, perhaps for the opposite reason \citep{Oishi2010}. 
However, in the balance and harmony module, Gallup was keen to explore the viability of a Likert framework, albeit one still adapted for their cross-cultural research context. 
First, items were framed in terms of time/frequency (i.e., how often people experience a given state) rather than size/amount (i.e., how much do they experience), as the latter might be somewhat abstract and harder for participants to envision. 
Second, a four-option scale was used (i.e., always, often, rarely, never) to minimize the complication with the frame. (Note: A three-option response scale was not viable, informed by Gallup's experience with people often choosing the middle category.)

In this study, we used a subset of data from the Gallup World Poll 2022 to illustrate how measurement properties (namely, thresholds) potentially differ for the Balance and Harmony module between the United States sample and the Norway sample.
However, a difficulty in comparing these data is that the response patterns tended to be skewed, leading to items with sparse response distributions and empty categories at the extremes.
We, therefore, used the proposed induced-Dirichlet prior structure while evaluating threshold posteriors while accounting for the uncertainty in estimation given the sparse response patterns.

A multiple-group item factor analysis was conducted using four different prior specifications.
The priors for thresholds were $ID(\bm\alpha=[1,1,1,1])$, $ID(\bm\alpha=[1,3,8,4])$, sequential exponential sum with small variance and large variance.
The second prior aligns more closely with expectations that responses are skewed with probabilities of 0.05, 0.20, 0.50, and 0.25.
The item factor model was identified by fixing the latent response distribution intercepts to zero and residual variance to one in each group.
The use of arbitrary identification constraints makes a direct comparison of the thresholds difficult \citep{Wu2016}.
We implemented alternative methods for identification in our Online Supplement for comparison but found that different threshold priors resulted in similar conclusions about how the posterior distributions would update. 
For brevity, we only discuss the more common fixed latent response constraints here.

\subsection*{Results}

\subsubsection*{Data Summary}
Norway and the United States are relatively comparable countries in terms of societal and economic development, and one might not generally expect any differences in the operating characteristics of the assessment tool across these two countries. 
The assessment of balance and harmony could reveal potential nuanced differences in the flourishing of these two countries. 
Indeed, it is intriguing that across the 8 items, if we take the percentage of respondents in each country who report experiencing the various aspects of balance and harmony either ``always'' or ``often,'' Norway does better overall (with higher levels of flourishing than the USA on 6 items), the USA nevertheless fares better on two of them. 
These include the amount of things happening in life is ``just right'' (Norway ranks 33rd out of 141 countries, with 72.7\% answering either ``always'' or ``often,'' versus USA at 25th with 74.3\%). 

However, such nuances aside, once we dive into these data, the Norway sample did not endorse \textit{Never} for the item ``Harmony Around'' (\textit{You Are in Harmony With Those Around You}; see Table \ref{tb:gwp-counts}).
No individuals in Norway endorsing \textit{Never} leads to a sparse response distribution.
If we estimated a multiple-group item factor model using a frequentist approach would lead to threshold 1 being unestimable.
As mentioned in the introduction, we could collapse the categories of \textit{Never} and \textit{Rarely} into one category to avoid a mismatch in the number of categories observed across groups.
The interested reader is referred to \citet{DiStefano2021} for a broader discussion on the effects of collapsing categories.
However, collapsing categories would limit our ability to generally compare the operating characteristics of each category across countries.

\begin{table}[!htp]
\centering
\begin{threeparttable}
\caption{Response distribution of balance and harmony items in Norway and US samples}
\label{tb:gwp-counts}
\begin{tabular}{lrrrrrrrr}
  \toprule
  & \multicolumn{4}{c}{Norway} & \multicolumn{4}{c}{United States}\\ \cmidrule(lr){2-5} \cmidrule(lr){6-9}
  Item & \textit{Never} & \textit{Rarely} & \textit{Often} & \textit{Always} & \textit{Never} & \textit{Rarely} & \textit{Often} & \textit{Always} \\ \midrule
Balance & 11 & 166 & 703 & 96 & 44 & 241 & 554 & 137\\
Amount Right & 28 & 237 & 601 & 110 & 58 & 194 & 561 & 163\\
Harmony Around & 0 & 36 & 753 & 187 & 14 & 71 & 673 & 218\\
Thoughts Harmony & 10 & 105 & 750 & 111 & 19 & 119 & 667 & 171\\
Stable & 7 & 50 & 470 & 449 & 21 & 85 & 520 & 350\\
Content & 8 & 85 & 754 & 129 & 31 & 101 & 592 & 252\\
Mind Ease & 17 & 163 & 655 & 141 & 52 & 182 & 611 & 131\\
Inner Peace & 27 & 234 & 610 & 105 & 29 & 172 & 584 & 191\\
Total Score & \multicolumn{4}{c}{24.0 (3.10)} & \multicolumn{4}{c}{23.9 (3.78)}\\
   \bottomrule
\end{tabular}
 \vspace*{1mm}
 	\begin{tablenotes}[para, flushleft]
    {\small
        \textit{Note.} These data represent the complete cases in each country subset to have an equal sample size. \textit{Never}=1, \textit{Rarely}=2, \textit{Often}=3, and \textit{Always}=4.
    }
 	\end{tablenotes}
 \end{threeparttable}
\end{table}

\subsubsection*{Item factor analysis-threshold estimates}

Comparing the measurement properties between these two countries using the original response scale (\textit{Never} to \textit{Always}) is preferred so as not to artificially induce a difference between countries \citep{Rutkowski2019}.
Using the methods developed earlier in this work, we estimated an item factor model for both groups to illustrate how thresholds are estimated within each country.
A comparison of thresholds between groups is not directly possible using the methods of this paper. 
The methods and comparisons of thresholds are illustrative of the influence prior specification for thresholds has on estimation in item factor analyses, and are not representative of an approach to invariance testing within a Bayesian framework for item factor analysis.
Using pairwise maximum likelihood (PML), the estimator of the threshold for the item ``Harmony Around'' does not exist for the Norway sample.
Next, we focus on comparing the posteriors when using different specifications of the induced-Dirichlet and the sequential sum with small or large variance.

\begin{figure}[!htp]
	\centering
	\includegraphics[width=0.65\textwidth]{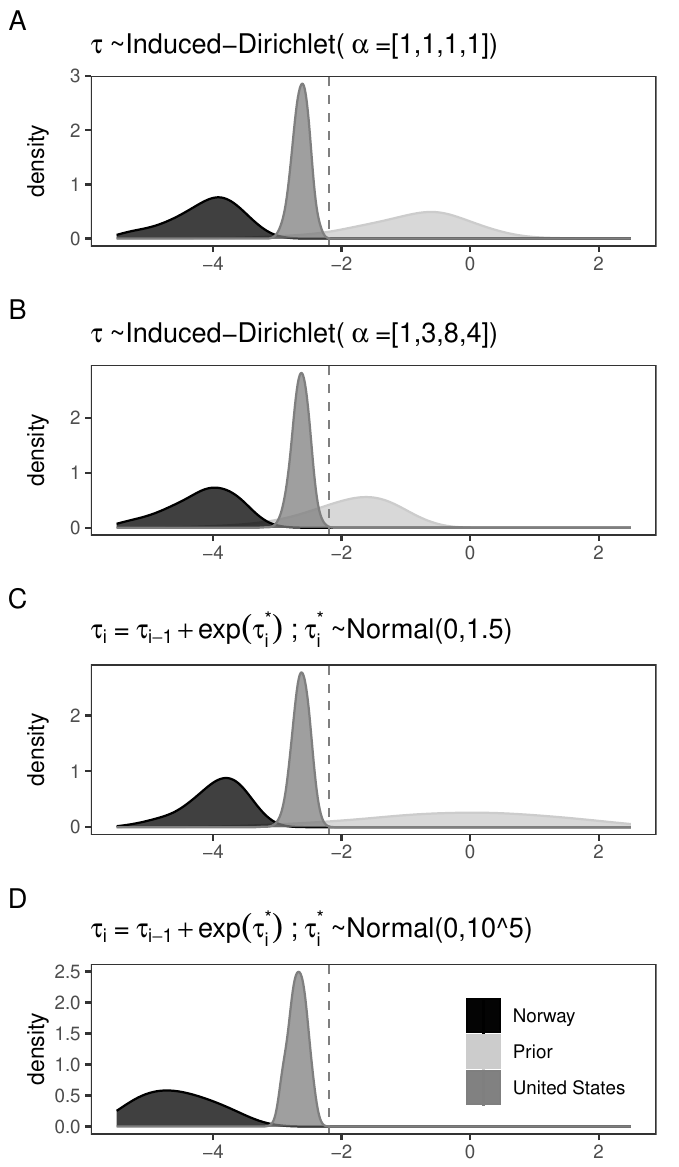}
	\caption{Comparison of \textit{Harmony Around} item threshold one posterior distributions among prior specifications}
	\label{fig:gwp-post-i3}
 \vspace*{-0.1cm}
\begin{flushleft}
\textit{Note}. Vertical (dashed) lines represent the estimated thresholds under pairwise maximum likelihood (PML in lavaan) for the United States. The threshold, tau[1], estimator does not exist for Norway under PML so a line is not present.
\end{flushleft}
\end{figure}

Of particular interest in this study are the thresholds for item 3 (\textit{Harmony Around}), where this Norway sample had no responses to the category ``Never.''
The first threshold posteriors for the Norway sample are generally lower than the US sample's except for items \textit{Stable} and \textit{Inner Peace} that were nearly identical.
The differences among these samples' estimated thresholds were less obvious for the second and third thresholds.
These results suggest that the individuals in the Norway sample were less likely to endorse the lowest response category compared to individuals in the US sample.
Of particular interest in this study was the difference in posterior distribution for the thresholds of item 3 (\textit{Harmony Around}), and the posteriors are shown in Figure \ref{fig:gwp-post-i3}.
We were particularly interested in how sensitive the difference in estimated thresholds was to prior specifications.
Four different prior specifications were used to show how sensitive the posteriors were to the choice of prior.
Using a default, relatively uninformative prior of $\alpha=[1,1,1,1]$ for the induced-Dirichlet specification resulted in posteriors that were nearly identical to the sequential exponential of normal distribution priors that is the default in blavaan.
In both specifications, we would conclude that the first threshold differs between Norway and the US.
If one modifies the induced-Dirichlet prior to $\alpha=[1,3,8,4]$, which supplies more weight to categories individuals are more likely to endorse.
However, the resulting posterior distributions were similar regardless of the prior specification.

\section*{Discussion}

The prevalence of questionnaires using ordered categorical responses in social and behavioral science calls for a robust understanding and set of tools for the most appropriate ways of analyzing such data. 
Although the importance of appropriate analysis has a statistical foundation, the ultimate goal of much, if not most, social science research is to make defensible and reasonable decisions about the constructs being evaluated.
Such decisions can affect the lives of individuals, families, communities, and society at large, so every effort must be taken to ensure that strong evidence supports the use of data collection instruments. Of course, the evidence supporting an instrument's use for a particular purpose can take many forms. 
Still, the evidence yielded by latent variable models, such as those described in this paper, is one such piece of evidence. 
The results of this paper will contribute to the extant literature on appropriate modeling techniques to support decision-making in social and behavioral science when researchers use categorical response options as part of their assessment procedure. 

The present analyses provide the first evaluation within a Bayesian latent variable modeling tradition of the effects of a joint threshold prior specification on parameter estimation, at least to the best of our knowledge.
Our results were partially consistent with results from a frequentist estimation tradition with respect to the effects of sparse response distribution on parameter recovery.
Our results aligned with the conclusions of bias of factor loadings and factor correlations \citep{DiStefano2014, DiStefano2021} when many items contain a sparse response distribution (e.g., a type of nonnormality).
Our results depart from prior conclusions that modeling sparse data leads to poor estimation performance \citep{Savalei2011, DiStefano2021}.
We found that the estimation performance under a Bayesian estimation procedure avoids, at least to a degree, the complex issues that arise when estimating polychoric correlations.
For instance, \citet{Savalei2011} demonstrated how parameter and standard error estimation deteriorates when a cell in a bivariate table of response categories is empty.
We saw some of these concerns with lower than nominal coverage rates of credible intervals when a sparse item was used as the indicator variable, but shifting to a different indicator variable significantly improved performance.
The use of Bayesian estimation with threshold priors that default to realistic values for the threshold appears to be a potentially viable solution.

The specification of priors for latent variable models is an ongoing area of research \citep{Depaoli2022, Miocevic2021, Zitzmann2020, Van_Erp2018}.
The attention of much of this research is on the specification of the priors for the latent variables or the latent regression parameters.
Similarly, the item response theory literature has a long history of the evaluation of priors for latent variables.
One aspect the broader latent variable modeling literature on prior specification can take away from the IRT literature is the use of joint priors for item parameters \citep[e.g., ][]{van_der_Linden2010}.
The proposed Dirichlet prior for threshold incorporates the joint prior specification for item parameters with the interpretative goals of priors for latent variables.

The intended interpretation of the Dirichlet prior for thresholds is as the number of ``pseudo''-observations endorsing each response option.
The straightforward interpretation provides a way to explain the results we observed in the Gallup World Poll Harmony module application.
In that example, the category ``Never'' was not endorsed for one item in a sample size of 976, but in a large enough sample, someone is likely to respond ``Never.'' 
The Dirichlet prior for thresholds provides a way to interpret how much additional information is incorporated into the analysis via the prior.
The proposed induced-Dirichlet prior specification helps in meaningfully specifying priors for the model and provides regularization in the estimation of thresholds when no endorsement is observed.
Regularizing the threshold estimates in such circumstances is needed because there is a nonzero probability of endorsing that category in the population of respondents.
Even if a response was not observed in a sample, a model that allows for that probability to exist, however small, is preferred because the model represents the population with more fidelity.

The regularizing effect of the induced-Dirichlet prior could be significant, even for relatively diffuse priors.
The use of the induced-Dirichlet prior might be enhanced by scaling the variance of the normal CDF within the change of variable phase based on how the latent response distribution is parameterized. 
For instance, this paper focused on the THETA parameterization leading to a latent response distribution with a larger-than-unit variance. 
If the induced-Dirichlet prior is relaxed such that the normal CDF incorporates the estimated latent response variance, the relative informativeness of the induced-Dirichlet prior would be reduced. 
This minor modification could be a valuable addition to make the induced-Dirichlet prior match the characteristics of the latent response distribution. 
An alternative could be the DELTA parameterization, where the latent response variances are fixed to 1. The advantage of this fixed variance is the natural connection with the induced-Dirichlet prior that utilizes the standard normal CDF.
However, using the DELTA parameterization comes with the complexity of specifying constrained priors for factor loadings and latent variable covariances.
Additional constraints on priors for loading and factor (co)variances would make specifying informative priors in applied examples more difficult.

The applied example of a multiple-group model highlighted the effects of prior specifications on posteriors.
Multiple group models are commonly used to help test measurement invariance; however, the methods demonstrated in this work are not representative of how to conduct invariance testing with Bayesian item factor analysis. 
The arbitrary identification constraints of the latent response distribution needed when estimating these models makes directly comparing thresholds between groups invalid under the methods used in this paper.
The methods in this paper are, though, potentially helpful in such invariance evaluations because previous research has shown how collapsing categories can lead to different conclusions about invariance, e.g., ``... collapsing also led to rejecting the assumption of slope equality in spite of a commensurate data-generating model.'' \citep{Rutkowski2019}. 
Utilizing the Dirichlet prior can provide a solution to not needing to collapse categories to investigate invariance but would instead need to rely on how threshold priors are specified to make conclusions about the degree of evidence in favor of or against invariance.
Conclusions would then be dependent on explicit assumptions about priors that can be evaluated instead of conclusions about invariance being dependent on how the categories were collapsed and, as Rutkowski and colleagues pointed out, could lead to issues of ``p-hacking'' and researcher degrees of freedom \citep{Simmons2011, Simonsohn2014}.

Evaluating the choice of the induced-Dirichlet prior may depend on whether the intended interpretation of pseudo-observations is reasonable.
At this point, we think interpreting the hyperparameter vector $\bm\alpha$ as pseudo-observations is appropriate because of the common use of this interpretation in simpler models.
For validation in more complex models, utilizing the delta-information prior approach may be a solution \citep{Morita2008, Morita2012}.
Evaluating the prior effective sample size would be valuable to identify if the interpretation of pseudo-observations is appropriate.

\subsection*{Recommendations}

The results of these studies highlight the importance of how observed data characteristics can negatively influence the estimation of Bayesian factor models.
When at least one item contains an empty response category, the estimates of all parameters are impacted to some degree.
We first recommend making sure not to use an item with an empty category as the reference indicator.
Secondly, when estimating a single group,  single time point factor model, differences were negligible between using a small variance sequential sum prior for thresholds versus the proposed joint induced-Dirichlet prior.
We found that using either approach will yield credible intervals with good coverage rates for most parameters under the conditions in this paper.

The similarity in results between the small variance sequential sum prior for thresholds and the joint induced-Dirichlet prior is promising that either could be a useful approach.
The sequential sum approach has the advantage of flexibility in adding constraints across items or groups for specific thresholds.
For example, a middle threshold could be constrained to equality across groups or items, and a free threshold parameter can be estimated within each group.
This would not be possible for the joint induced-Dirichlet prior because all thresholds within an item would need to be constrained.
This limitation of the induced-Dirichlet prior might be able to be overcome by manipulation of the hyper-parameter vector ($\bm\alpha$), but how this could be accomplished is an open question.
Allowing for constraints between groups is an important consideration when moving into multi-group factor models.

Comparing threshold parameters between groups is less certain when the number of thresholds varies because of empty categories.
The example applied analysis showed how the uncertainty of observing an empty category can be taken into account without restricting the range of the observed data to force an equal number of categories.
However, this is not yet easily implemented in blavaan or Mplus.
We think, based on the results of this paper, such an approach can more accurately model categorical data with latent variable models without modifying the characteristics of the observed data.

\subsection*{Conclusion}

Modeling messy data can require difficult analytic decisions, including transforming the observed data or developing more complex methods to account for the messiness.
We demonstrated an approach to estimating thresholds in Bayesian item factor analysis that accounts for the inherent messiness in data commonly encountered in social science research.
Researchers can utilize the regularizing joint prior on thresholds to account for issues with small sample sizes or heavily skewed response distributions to better estimate their models with confidence that their results are less influenced by the characteristics of messy data.

\subsection*{Acknowledgments}
The authors would like to thank Roy Levy from Arizona State University for helpful comments on an earlier draft of this paper presented at the American Educational Research Association annual meeting in 2023.
Many of the computations in this paper (especially Study 2 and Study 3) were run on the FASRC Cannon cluster supported by the FAS Division of Science Research Computing Group at Harvard University.

\subsection*{Data Availability Statement}
Additional results and details about the analyses and data presented in this manuscript are available in our Online Supplement and in our accompanying online repository (\url{https://osf.io/kpvwb/}). The code is openly available, and most data in this paper were simulated. The Gallup World Poll data are available by request at \url{https://www.gallup.com/analytics/468179/global-wellbeing-initiative-dataset.aspx}.

\bibliographystyle{apalike}
\bibliography{bayes_threshold_paper}

\section{Sampling Latent Response Distribution}\label{app:stan}

The following Stan code is based on the implementation of \citet{Goodrich2017}.

{
\footnotesize
\begin{lstlisting}
  //y = observed response vector
  //mu = latent response mean vector (if marginal likelihood, then this is 0
  //L = lower Cholesky decomposition of covariance matrix of latent responses
  //b = matrix (nitems by nlevs) of thresholds
  //u = nuisance (0,1) variable
  //nlevs = vector designating number of levels/number of categories for each item
  vector[] tmvn(int[] y, vector mu, matrix L, vector[] b, real[] u, int[] nlevs) {
    int K = rows(mu);
    vector[K] d;
    vector[K] z;
    vector[K] out[2];
    for (k in 1:K) {
      int km1 = k - 1;
      real nu;
      if (y[k] == 1) {
        real z_star = (b[k,1] - (mu[k] + ((k > 1) ? L[k,1:km1] * head(z, km1) : 0))) /  L[k,k];
        real u_star = Phi(z_star); //normal CDF for implied density of TMVN
        nu = u_star * u[k];
        d[k] = u_star;
      } else if(y[k] == nlevs[k]) {
        real z_star = (b[k,(nlevs[k]-1)] - (mu[k] + ((k > 1) ? L[k,1:km1] * head(z, km1) : 0))) /  L[k,k];
        real u_star = Phi(z_star);
        d[k] = 1 - u_star;
        nu = u_star + d[k] * u[k];
      } else {
        int lb = y[k]-1;
        int ub = y[k];
        real z_starL = (b[k,lb] - (mu[k] + ((k > 1) ? L[k,1:km1] * head(z, km1) : 0))) / L[k,k];
        real z_starU = (b[k,ub] - (mu[k] + ((k > 1) ? L[k,1:km1] * head(z, km1) : 0))) / L[k,k];
        real u_starL = Phi(z_starL);
        real u_starU = Phi(z_starU);
        nu = u_starL + (u_starU - u_starL)*u[k];
        d[k] = u_starU - u_starL;
      }
      z[k] = inv_Phi(nu); //convert back to z-score from uniform variate
    }
    out[1] = z; //simulated ystar value
    out[2] = d; //density
    return(out);
  }
\end{lstlisting}
}
\section{Deriving informative prior for sequentially defined thresholds} \label{app:sen}

Placing informative priors on individuals thresholds is not entirely obvious.
For instance, consider the situation with an assessment where items are scored with four ordered categories, and previous results are available to create informative priors for the thresholds.
The reported threshold point estimates for one of the items are $-$2.00, $-$0.25, and 1.75.
An informative prior for the first thresholds is straightforward, 
$$\tau_1 \sim \text{Normal}(-2.00, 0.20).$$
An informative prior for the second threshold is unfortunately not simply $\tau_2 \sim \text{Normal}(-0.25, 0.25)$, but is instead specified \textit{relative} to threshold 1. 
The informative prior for $\tau_2$ needs to be constructed using the solution to the following system.
\begin{enumerate}
\item We need to identify the distribution of $\tau^{\ast}_2$ such that $\tau_2 \sim \text{Normal}(-0.25, 0.25)$.
\item Rewrite $\tau_2 = \tau_1 + exp(\tau^{\ast}_2)$ in terms of the expected value and variances of each element, taking advantage of the fact that $exp(\tau^{\ast}_2) \sim \text{logNormal}(.)$.
\begin{align*}
E[\tau_2] &= E[\tau_1] + exp\left(E[\tau^{\ast}_2] + \frac{Var[\tau^{\ast}_2]}{2}\right)\\
Var[\tau_2] &= Var[\tau_1] + \left(exp\left(Var[\tau^{\ast}]\right)-1\right)exp\left(2E[\tau^{\ast}_2] + Var[\tau^{\ast}_2]\right),\ \tau_1 \perp \tau^\ast_2,
\end{align*}
\item The prior we need to specify to obtain the desired $E[\tau_2]$ and $Var[\tau_2]$ is then the solution system of equations above for values of $E[\tau^{\ast}_2]$ and $Var[\tau^{\ast}_2]$ under the conditions that $E[\tau_1]<E[\tau_2]$ and $Var[\tau_1] < Var[\tau_2]$. The solution is not immediately obvious but is analytically defined.
\item The solution (after some algebra) is  
\begin{align*}
E[\tau_c] &> E[\tau_{c-1}]\\
Var[\tau_c] &> Var[\tau_{c-1}]\\
E[\tau^\ast_c] &= log\left(E[\tau_c] - E[\tau_{c-1}]\right) - \frac{Var[\tau^\ast_c]}{2}\\
Var[\tau^\ast_c] &= log\left(\frac{Var[\tau_c] - Var[\tau_{c-1}] + exp\left(2log( E[\tau_c] - E[\tau_{c-1}] )\right)}{exp\left(2log(E[\tau_c] - E[\tau_{c-1}] )\right)}\right).
\end{align*}
\item The resulting prior for $\tau^\ast_2$ one needs to use to obtain the prior we want to place on $\tau_2$ is 
$$\tau^\ast_2 \sim \text{Normal}(0.55, 0.02).$$
\end{enumerate}
A major caveat is that the resulting prior on $\tau_2$ is only approximately normally distributed (the larger the variance of $\tau^\ast_2$ is, the more asymmetric the prior on $\tau_2$ will be), though we will have the desired expected value and variance.

\section{Induced-Dirichlet Prior Implementation} \label{app:thresh}

The Dirichlet distribution is defined as
\begin{equation}
\text{Dirichlet}(\theta\vert \alpha) = \frac{\Gamma\left(\sum_{k=1}^K\alpha_k\right)}{\prod_{k=1}^k\Gamma\left(\alpha_k\right)}\prod_{k=1}^{K}\theta_k^{\alpha_{k}-1},
\end{equation}
where $K\in \mathbb{N}, \alpha\in(\mathbb{R}^{+})^K$, for a $\theta\in K$-simplex.
The hyper-parameter vector $\bm\alpha$ is strictly positive, with each element $\alpha>0$.
Sampling from the Dirichlet distribution produces a vector of probabilities that sum to unity.
The probability density function for the Dirichlet distribution can then be used to compute the likelihood of a given vector of probabilities that define a simplex based on the given hyper-parameter for the Dirichlet distribution.
In this work, the Dirichlet distribution is used to compute the probability/likelihood associated with a given simplex of probabilities induced by a set of thresholds for an item.

The density function of the Dirichlet distribution provides a convenient mechanism for using the probabilities induced by a set of thresholds for an item. 
We can randomly sample thresholds \textit{indirectly} through the Dirichlet distribution.
First, the simplex of probabilities are simulated, then thresholds are computed based on the simulated probabilities.

{
\begin{lstlisting}
  vector phi_logit_approx_vec(vector b) {
    int N = num_elements(b);
    vector[N] a;
    a = inv_logit( 1.702*b );
    return(a);
  }
  //c = vector of thresholds
  //alpha = dirichlet hyper-prior parameters default=rep(1, num_elements(c)+1)
  //phi = center of latent response distribution, default=0
  real induced_dirichlet_lpdf(vector c, vector alpha, real phi) {
    int K = num_elements(c) + 1;
    vector[K - 1] anchoredcutoffs = c - phi; //which is first is important c vs. phi...
    vector[K] sigma;
    vector[K] p;
    matrix[K, K] J = rep_matrix(0, K, K);
    sigma[1:(K-1)] = phi_logit_approx_vec(anchoredcutoffs);
    sigma[K] = 1;
    p[1] =  sigma[1];
    for (k in 2:(K - 1))
      p[k] = sigma[k] - sigma[k - 1];
    p[K] = 1 - sigma[K - 1];
    for (k in 1:K) J[k, 1] = 1;
    for (k in 2:K) {
      real rho = 1.702 * sigma[k - 1] * (1 - sigma[k - 1]);
      J[k, k] =  - rho;
      J[k - 1, k] = + rho;
    }
    return dirichlet_lpdf(p | alpha)
            + log_determinant(J);
  }
\end{lstlisting}
}

\end{document}